# The Structure of Scientific Socialism: Quantum Emergence, Frustration, and the Non-Dual Dialectic

Sindhunil Barman Roy

Department of Physics, School of Mathematical Science

Ramakrishna Mission Vivekananda Educational and Research Institute

Belur Math, Howrah

India

And

Department of Physics

Indian Institute of Technology Bombay

Powai, Mumbai - 400076

India

Email:sindhunilbroy@gmail.com

**Abstract:** Classical Marxism and the "algebra of revolution" were formulated within the ontological constraints of 19th-century Newtonian materialism—a world of discrete, predictable, "billiard-ball" interactions. However, the 20th-century transitions in physics, from Thomas Kuhn's paradigm shifts to Phil Anderson's philosophy of emergence, have dismantled the reductionist foundations of this mechanical worldview. This paper proposes a "New Manifesto" for Scientific Socialism by synthesizing modern condensed matter physics with the non-dual philosophy of Advaita Vedanta. By examining the concepts of "Geometric Frustration" and "Competing Interactions" through the lens of Spin-Glasses and Mott Insulators, we argue that social "stasis" and "synthesis" are emergent properties of a universal consciousness field rather than mechanical inevitabilities. We further explore how this quantum-informed dialectic resolves the "essential tension" between the individual and the collective, echoing the intuitions of Schrödinger and Heisenberg regarding the foundational unity of reality.

# I. INTRODUCTION: The Meltdown of Mechanical Materialism

The evolution of human society has long been viewed through the lens of objective transitions, but as our understanding of the physical world shifts from the clockwork precision of Newton to the probabilistic landscape of quantum science, our social philosophies must follow suit. The *"Scientific Socialism"* of the 19th century was an *"algebra of revolution"*—a term famously coined by Alexander Herzen to describe the Hegelian dialectical method as a symbolic framework for solving for social unknowns. While Karl Marx and Friedrich Engels successfully grounded this algebra in the material world, they did so under the paradigm of *"Mechanical Materialism,"* where matter consisted of isolated, objective parts governed by fixed, predictable laws (Marx, 1852).

However, the early 20th century triggered what Thomas Kuhn described as an *"Essential Tension"* (Kuhn, 1962,1977)—an accumulation of anomalies that the Newtonian paradigm could no longer resolve, leading to a total structural collapse of the old worldview. The resulting Quantum Revolution did not merely add new facts; it melted the very foundation of objectivity. The "objective" observer was replaced by the active participant, and the solid "billiard balls" of matter evaporated into Schrödinger wavefunctions—clouds of probability where particles are instantaneously connected across space, defying the localized logic of classical materialism.

This shift in physical ontology has naturally precipitated a similar movement in the social sciences. *Quantum Social Science (QSS)* (Höne, 2017) has emerged as a diverse research area that challenges the classical materialist assumption that social life and consciousness are purely classical, deterministic phenomena. QSS does not represent a unified scholarly body as yet, but encompasses various trajectories, from using quantum probability to model human decision-making to proposing that the world itself is fundamentally quantum. Pioneers like *Alexander Wendt* have argued for a radical "quantum realism," suggesting that human subjects can be rethought as "quantum walking waves" (Zonotti, 2025).

While the field of QSS has gained significant traction, existing literature often focuses on cognitive modelling or high-level political theory. For instance, *Holtfort and Horsch (2023)* utilize quantum decision theory to explain cognitive biases and Darwinian selection, yet their focus remains primarily on behavioural economics and evolutionary processes. Similarly, *Zanotti (2025)* explores quantum theory as an ontological critique to rethink ethics and relationality in international politics, moving away from Newtonian mechanistic models to embrace uncertainty and entanglement.

The present work distinguishes itself by grounding the social dialectic in the specific physical phenomena of *Condensed Matter Physics*. We contend that the core engine of the dialectic—the struggle between opposing forces—has evolved into a sophisticated modern avatar: *Frustration.* In modern physics, frustration—arising from geometric constraints or random competing interactions—prevents a system from settling into a static state, giving rise to "exotic" possibilities like *Spin-Glasses and Mott Insulators*. These phenomena provide a rigorous social mirror for the complexities of human history, where internal contradictions and "traumatized" memories of past revolutions prevent a simple return to equilibrium.

Furthermore, we align this physical framework with the *"Philosophy of Emergence"* championed by physics Nobel Laureate *Phil Anderson*, whose mantra "More is Different" dismantled the reductionist dream of understanding the whole simply by dissecting the parts (Anderson, 1972). Socialism, in this light, is reimagined not as a mechanical program to be imposed, but as an emergent property of a complex, interacting human "Quantum" field.

Finally, we explore the *"Non-Dual Bridge."* As quantum science pervades our daily technology, it increasingly correlates with the ancient Indian knowledge system of *Advaita Vedanta*. Building on the very recent theoretical framework proposed by *Strømme (2025)*, which posits universal consciousness as a foundational field rather than an emergent property of neural processes, we argue that the perceived separation between individuals is a *"functional illusion" (Maya)*. While Strømme and others draw from various metaphysical traditions, they often overlook the rigorous, thousand-year-old framework of Advaita Vedanta, which *Swami Vivekananda* succinctly expounded for the modern audience (Vivekananda, 1893, 1897, 1947, 1994). This synthesis provides the ultimate *"Algebra of Revolution"* for the 21st century: a Scientific Socialism where the *"Material"* is recognized as an interconnected field, and "Socialism" is the natural expression of a humanity that has realized its fundamental unity.

## II. The Algebra of Revolution and the Newtonian Constraint

The foundational logic of *"Scientific Socialism"* was built upon what Alexander Herzen famously termed the *"Algebra of Revolution"* (Rees, 1998). For Herzen, and subsequently for the Marxist tradition, the Hegelian dialectic served as a symbolic framework that allowed for the navigation of social upheaval without requiring the pre-definition of every specific historical variable. By applying a "scientific" prefix to socialism, Marx and Engels signalled a radical commitment to a materialist ontology, one that sought to rescue the dialectic from its "mystical shell" in Hegelian idealism and ground it in the physical reality of production and class struggle.

However, this 19th-century materialism was inextricably tethered to the dominant scientific paradigm of its era: *Newtonian Mechanics*. In the Newtonian universe, reality is a predictable, clockwork machine. Matter is composed of discrete, "billiard-ball" atoms governed by linear, reversible forces. This "Mechanical Materialism" permeated early Marxist thought, leading to a socio-political determinism where the transition from Capitalism to Socialism was often characterized as a mathematical inevitability (Spirkin, 1983, Roy, 1940).

Within this constraint, the relationship between the *Economic Base* and the *Legal-Ideological Superstructure* was frequently interpreted through a lens of strict causality. It was believed that if one could map the initial conditions of the forces of production—much like calculating the trajectory of a projectile—the future state of the superstructure could be predicted with Newtonian certainty. This left little room for the non-linear "ruptures" or the emergent complexities that characterize modern social systems.

As the 20th century unfolded, this mechanical view faced what Thomas Kuhn (Kuhn, 1962, 1977) identified as the *"Essential Tension"*. Kuhn argued that scientific progress does not occur through a smooth, linear accumulation of facts, but through the accumulation of *anomalies*—phenomena that the existing paradigm (in this case, mechanical materialism) cannot explain. When the weight of these anomalies becomes too great, the system enters a state of crisis, eventually triggering a *Paradigm Shift*.

In the physical world, this shift was the Quantum Revolution. In the social world, the anomaly was the persistence of systems that refused to follow the "Newtonian" trajectory of class collapse. The transition from the certainties of Newton to the probabilistic, entangled reality of the quantum meant that the "material" in materialism had effectively dissolved.

We must now update the *"Algebra of Revolution"* to account for a post-Newtonian ontology. In this new framework:

- *The Observer and the Observed*: Just as the act of measurement collapses the wavefunction in physics, the conscious "observer" (the social actor) is inseparable from the "observed" (the social system). Human agency is not a mere byproduct of the economic base; it is an active participant in the collapse of social probabilities.

- *Non-Linearity:* Social change is no longer viewed as a steady, predictable march. It is characterized by *Quantum Leaps*—sudden, structural shifts born from the *"frustration"* of competing interactions that do not follow a linear path of least resistance.

- *Entanglement*: Social forces are not isolated *"billiard balls"* bumping into one another. They are non-locally entangled. A shift in the *"Quantum Field"* of global consciousness can trigger instantaneous ripples across the social superstructure, defying the localized causality of 19th-century thought.

By breaking the Newtonian constraint, Scientific Socialism evolves from a rigid, deterministic prophecy into a dynamic, emergent science of complexity.

## III. Frustration and the Dialectic of Exotic Matter

The core engine of the *dialectic*—the struggle between opposing forces—has found a sophisticated modern avatar in condensed matter physics: *Frustration*. In a classical, non-frustrated magnetic system, individual spins (the *"actors"* of the system) easily align to minimize total energy, reaching a state of peaceful equilibrium. However, *"Frustration"* occurs when the system's internal architecture or the nature of its interactions prevents it from satisfying all its competing forces simultaneously.

This phenomenon arises from two distinct but often overlapping sources. First, *Geometric Frustration* (Moessner and Ramirez, 2006), where the lattice arrangement (such as a triangular or Kagome lattice) makes it mathematically impossible for every neighbouring spin to align in a way that satisfies its

neighbours—a physical "essential tension." Second, *Random Competing Interactions*, where the forces themselves are in direct contradiction (e.g., some favouring alignment, others favouring opposition) (Binder and Young, 1986). This dual-natured frustration prevents the system from settling into a static, "billiard-ball" equilibrium, giving rise to "exotic" states of matter that serve as a rigorous social mirror for the dialectics of class struggle.

In a standard ferromagnet, the *"Thesis"* (up-spin) and *"Antithesis"* (down-spin) resolve quickly into a uniform, low-energy state. In a *Spin-Glass*, however, the interactions are so conflicting and frustrated that the system possesses a *"complex energy landscape"* with nearly infinite "local minima" or possible states (Binder and Young, 1986, Anderson, 1990, Mydosh, 2015).

Crucially, spin-glasses exhibit *"memory" and "aging" effects*. If the system is disturbed, it does not return to its original state; it "remembers" its path. Much like a society shaped—or traumatized—by its past revolutions, a spin-glass cannot simply "reset" to a blank slate. Its current identity is an emergent product of its historical struggle. This is the physical realization of the Marxist observation that "the tradition of all dead generations weighs like a nightmare on the brains of the living" (Marx, 1852). In this state, the dialectic is not "solved," but is instead "locked" into a complex, multi-stable history.

Classical band theory predicts that any material with a partially filled electron band should be a metal—a conductor where electrons flow freely. Yet, in a *Mott Insulator*, intense electron-electron repulsion (Coulomb interaction) creates a state of quantum "frustration" that locks the electrons in place (Imada et al., 1998, Roy, 2019).

The classic Mott Insulator is often linked with a *"frustrated magnetism"* (Leiner 2019). It represents a profound state of social "stasis," where internal class or structural contradictions—the "repulsion" between competing interests—prevent the "flow" of social progress (the current) that the system's theoretical resources would otherwise suggest. The transition from a Mott Insulator to a metal—the *Mott Transition*—is not a gradual change but a sharp, paradigm-shifting leap. It serves as a physical model for the revolutionary *"Synthesis,"* where a qualitative change in the "density" of social interactions suddenly unlocks the potential of the entire collective.

In *quantum topological phases* of matter, the system's fundamental properties are not determined by individual, localized particles, but by the

*Global Structure (topology)* of the entire quantum state (Wen, 2017). Local disturbances, such as the removal or "breaking" of a single part, do not affect the stability of the whole because the information is stored non-locally.

This represents the ultimate physical realization of the dialectical *"Synthesis"*: a state where the collective "Whole" protects and dictates the reality of the "Parts." In a topological society, the "essential tension" between the individual and the collective is resolved through a global symmetry that provides a stability far greater than the sum of its isolated members. This move from local "frustration" to global "protection" mirrors the transition from a fragmented capitalist society to a unified, scientific-socialist framework.

## IV. The Philosophy of Emergence: "More is Different" as a Dialectical Bridge

As the 20th century progressed, the final pillar of the Newtonian social model—Reductionism—began to collapse under the weight of condensed matter physics. Reductionism, the epistemological dream that a system can be entirely understood by dissecting it into its smallest constituent parts, provided the "billiard-ball" logic for early mechanical materialism. It suggested that if we understood the individual worker or the single commodity, we could, through simple addition, understand the global economy. This dream was dismantled by Nobel Laureate Phil Anderson in his seminal 1972 paper, "More is Different" (Anderson, 1972).

Anderson's thesis provides a rigorous physical foundation for the Marxist law of the transition from quantity to quality. He argued that the ability to reduce everything to simple fundamental laws does not imply the ability to start from those laws and reconstruct the universe (Anderson, 1972). At each level of complexity, entirely new laws, symmetries, and properties appear that are not merely "more" of the same, but are fundamentally different in kind—a process known as Emergence.

In physics, emergence often occurs through the mechanism of "Broken Symmetry." Consider a collection of atoms in a liquid state: it possesses high symmetry because the atoms are randomly distributed and every direction looks the same. However, as the system reaches a critical threshold, it undergoes a phase transition into a crystal. In this "stiff" structure, the continuous translation symmetry is broken; atoms are now locked into specific lattice points. This

"broken symmetry" gives birth to entirely new macroscopic properties, such as rigidity and Bragg diffraction, which were non-existent and unpredictable at the liquid level (Chaikin and Lubensky, 1994).

Building on the work of Nobel Laureate Phil Anderson, the hierarchical structure of science demonstrates that *"More is Different"*. As systems grow in complexity, they transition through various levels—from quantum electrodynamics and chemistry to molecular biology, cell biology, and eventually to populations and society. Each level depends on the lower-complexity levels beneath it, but the behaviour at a higher level typically cannot be understood solely in terms of the underlying dynamics of the lower level. Consequently, there is no single, universally appropriate modelling framework that applies across all these scales. Instead, distinct theoretical frameworks are required at different stages of complexity, as illustrated in the Figure 1.

In this framework (see Figure 1), while quantum mechanics and quantum field theory are foundational for describing elementary particles and atoms, they eventually reach a limit of applicability as complexity increases. This separation between levels implies that knowledge at a lower level—such as the relevance of quantum field theory—is no longer fully informative about the emerging behaviour at higher levels like developmental biology or culture (Stumpf, 2022).

Recent interdisciplinary studies have expanded this concept into the social realm. For example, Holtfort and Horsch (2023) demonstrate how quantum-like decision-making processes lead to emergent "biases" and Darwinian selection that cannot be explained by classical utility maximization alone. Furthermore, the role of *"Frustration"*—both geometric and interactional—acts as a catalyst for these transitions. As discussed in the context of Kinetic Arrest (Chattopadhyay 2005, Roy and Chaddah, 2014) of a first order phase transition, systems under high frustration do not merely fail; they enter "arrested" states or reorganize into exotic symmetries.

Applying this to the Base-Superstructure relationship, the Superstructure (culture, law, and ideology) is an emergent level of reality. The collective interactions and the *"Frustration"* within the Base reach a critical point—a Criticality—where a new social symmetry is born. This is the qualitative leap that Thomas Kuhn (Kuhn,1962) identified as a Paradigm Shift, where the old rules of "Normal Science" are discarded for a new organizational logic.

In the classical "Newtonian-Marxian" view, socialism was often framed as a mechanical inevitability—the logical "output" of a failing capitalist "input." Anderson's philosophy of emergence suggests a more dynamic, non-linear alternative. Socialism, in this light, is an Emergent Property of a complex, interacting human "quantum" field.

It is a higher-order organization that appears when the "Essential Tension" and internal "Frustration" of the capitalist paradigm can no longer be contained within the old symmetry. When the quantitative contradictions of capital reach a tipping point, the system re-organizes. This is the ultimate physical realization of the Synthesis: a state that is fundamentally "different" in its governing laws. In an emergent socialist framework, the laws of the "Whole" (the collective) dominate the properties of the "Parts" (the individuals), creating a state of topological stability impossible under fragmented, reductionist logic.

However, the paradigm of emergence meets a significant boundary when applied to the most complex phenomena: life and consciousness. As Kivelson and Kivelson (2016) argue in their *npj Quantum Materials* perspective, defining emergence strictly within the "thermodynamic limit"—where the number of constituents ($N$) approaches infinity—creates a "peculiar dilemma."

If life and consciousness are only sharply defined in this infinite limit, then biological entities are only "approximately alive and operationally conscious." This reveals the danger of extending a purely physical definition of emergence into the realm of the animate. It suggests a fundamental gap: emergence explains the *reorganization* of matter, but it struggles to account for the foundational "spark" of consciousness itself.

This "Kivelsons Dilemma" is precisely where the *Non-Dual Bridge* becomes necessary. If we are to move beyond being "operationally conscious" to being fundamentally unified, we must look beyond the physical interactions of $N$ particles. We must turn to the *Advaita Vedanta* perspective, where the "Whole" (Brahman) is not a derivative byproduct of infinite parts, but the *primary, foundational reality*. In this view, the "Synthesis" is not merely an emergent reorganization of matter, but a realization of the foundational field of consciousness that precedes all differentiation.

## V. The Non-Dual Bridge: Quantum Science and Advaita Vedanta

The "restless teenager" of 21st-century quantum science has matured beyond the laboratory, pervading our daily technology from the ubiquitous transistors in mobile phones to the advanced hardware of *Spin-Glass* and *Topological materials*. However, as we push into the ontological foundations of this quantum reality, we find that the "High Priests" of modern physics—Schrödinger, Heisenberg, and Bohr—were forced to look toward the ancient Indian knowledge system of *Advaita Vedanta* to find a logic capable of housing quantum paradoxes (Moore, 1989).

The hallmark of the quantum world is *Entanglement*, a phenomenon where particles that have interacted remain part of a single, unified state regardless of the spatial distance between them. This fundamentally shatters the classical, reductionist idea of "locality"—the notion that an object is only influenced by its immediate surroundings. In the non-dual philosophy of Advaita, this "separation" between individuals or objects is defined as *Maya*, a functional illusion that masks the underlying unity of existence.

Very recent theoretical frameworks of Strømme presented in *AIP Advances* (Strømme 2025), have proposed that consciousness itself may be a foundational field, with "Mind" and "Thought" acting as the drivers of differentiation within that field. While that specific paper explores various modern non-dual sources, it curiously overlooks the thousand-year-old rigorous philosophical framework of *Advaita Vedanta*. This is a significant omission, as it was *Swami Vivekananda* who, in the late 19th century, succinctly expounded this philosophy for the modern scientific audience, bridging the gap between ancient intuition and the requirements of empirical reason.

This synthesis provides the scientific-socialist with a profound new *"Algebra of Revolution"*:

- *The Thesis:* The Individual (The Classical Illusion of Separation).

- *The Antithesis:* The Collective (The Frustrated Interaction and the struggle for localized equilibrium).

- *The Synthesis:* The Non-Dual Field (The Realization of Oneness/Brahman through the collapse of the localized ego).

The *"Structure of Scientific Socialism"* for the 21st century is not found in the mechanical re-distribution of goods, but in the synthesis of *Quantum*

*Science and Advaita Vedanta*. By recognizing that the "Material" is an interconnected field and that "Socialism" is the emergent expression of our shared oneness, we move beyond the gatekeeping of the old "High Priests"—both the ancient clerical elites and the modern academic "Neo-Macaulayites."

As Swami Vivekananda (1897) argued in his lectures from *Colombo to Almora*, the realization of the "Self" in the "Other" is the only stable foundation for a truly just society. When the "Other" is recognized as a part of the same non-dual field, the exploitation inherent in the capitalist paradigm becomes an act of self-harm. In this light, Socialism is revealed as the macroscopic manifestation of the microscopic reality of entanglement. It is the social state that corresponds to the highest physical and philosophical truth: that at the fundamental level, the field is one.

## VI. Advaita Vedanta: Philosophy, Civilization, and the Critique of Religion

It is necessary here to address the potential tension between Marxian materialism and the non-dual traditions of the East. When Karl Marx (Marx 1844) famously characterized religion as the *"opium of the people,"* his critique was specifically situated within the socio-historical context of 19th-century European Abrahamic structures. In that framework, religion functioned as a vertical, top-down mechanism—a "single superpower" or an external Godhead—that justified the existing class hierarchy and provided a "heart" to a heartless world, albeit one that pacified the revolutionary impulse.

However, applying this critique to *Advaita Vedanta* constitutes a category error. To understand the synthesis proposed in this paper, two critical distinctions must be made.

First, Advaita Vedanta is not a "religion" in the sectarian or dogmatic sense; it is a rigorous philosophical system among several variants of ancient Indian thought. Its foundations are found in the *Shrutis*, which, rather than being static "holy books," functioned more akin to modern *academic journals*—a decentralized collection of insights, debates, and empirical-subjective observations recorded over centuries.

While *Adi Shankaracharya* (c. 8th century CE) codified these insights into a structured "textbook" form, it was Swami Vivekananda in the late 19th century who "democratized" the philosophy. Swami Vivekananda (Vivekananda

1893,1897) stripped away the esoteric barriers, making the logic of non-dualism accessible to a global audience and demonstrating that its principles were compatible with—and indeed anticipatory of—the laws of modern science. As an ontology of the *"Field,"* Advaita Vedanta is concerned with the fundamental nature of reality, not the appeasement of a celestial monarch.

Secondly, it is historically and sociologically inaccurate to bracket the Indian tradition—often broadly termed *"Hinduism"*—within the same categorical framework as Abrahamic religions. As Swami Vivekananda frequently articulated (Vivekananda 1893,1894,1937,1947,1994), what the West calls *"Hinduism"* is not a centralized religion of a single book, prophet, or creed; it is a *Civilization*.

Unlike the Abrahamic model, which is often defined by "Belief" and the maintenance of a vertical relationship with a creator, the Indian civilizational model is defined by *Inquiry* and the realization of horizontal unity. Vivekananda's distinction is vital for the modern socialist: if the goal of socialism is the dissolution of the *"Other"* and the realization of a collective whole, then *Advaita Vedanta* provides the philosophical hardware for that realization without the *"opium"* of a pacifying, external deity. In this light, the *non-dual field* is not a religious *"heaven"* but a scientific and social *"truth"* that demands active engagement and the dismantling of the illusory barriers of the ego.

## VII. Conclusion: Breaking the Epistemic Gatekeepers

The primary obstacle to the realization of a Scientific Socialism in the 21st century is the *"Essential Tension"* maintained by modern institutional gatekeepers. As this paper has demonstrated, the perceived complexity of higher mathematics and the purported "mysteries" of quantum physics are frequently utilized as tools of social and intellectual exclusion. Historically, educational systems, especially in developing world, designed under colonial or Eurocentric frameworks sought to bifurcate "Science" from "Heritage," a division that continues to haunt modern academic structures.

This gatekeeping is most visible in the asymmetrical valuation of non-dual philosophy. While renowned American scholars like Diana Eck at Harvard spend decades in rigorous engagement with Indian non-dual traditions—extracting the "cream" of this knowledge to synthesize a more profound theological and pluralistic understanding of the global landscape (Eck, 2012)—many modern sociologists within the developing world remain trapped in a reflexive, post-colonial scorn for their own intellectual heritage.

This creates an academic *"Neo-Colonialism"* where the tools of liberation—both the *Algebra of Marx* and the *Non-Dual Logic of Vivekananda*—are rendered opaque and inaccessible to the general populace. The democratization of these concepts is essential; the "mystery" must be stripped away to reveal that the logic of the quantum vacuum and the logic of the non-dual field are one and the same.

The global audience must recognize that the rapid advancement of *Artificial Intelligence* and *Quantum Hardware* are not anomalous "monsters" of a runaway technocracy, but the natural, predictable consequences of the *Principle of Emergence*. These technologies are the physical manifestations of the "More is Different" philosophy, representing a transition to a higher level of complexity in human interaction.

Intellectual sovereignty in this new era does not require a retreat into an archaic past or a decade-long immersion in the labyrinth of ancient Puranic texts. Rather, it requires a focused engagement with the modern synthesis provided by *Swami Vivekananda* (Vivekananda 1897, 1947, 1994). In his work, the synthesis is effectively finalized: the individual is understood as the field, the social struggle is identified as the physical frustration of competing interactions, and the revolution is defined not as a mechanical collapse, but as the qualitative leap into a unified, non-dual state of existence.

By bridging *Condensed Matter Physics, Marxian Dialectics,* and *Advaita Vedanta*, we move beyond the mechanical materialism of the 19th century toward a truly *"Scientific Socialism"*. This is a system where the "Material" is no longer a collection of isolated billiard balls, but an interconnected, entangled field. The algebra of revolution is thus evolved: the transition to a just society is the macroscopic emergent expression of the microscopic reality of oneness.

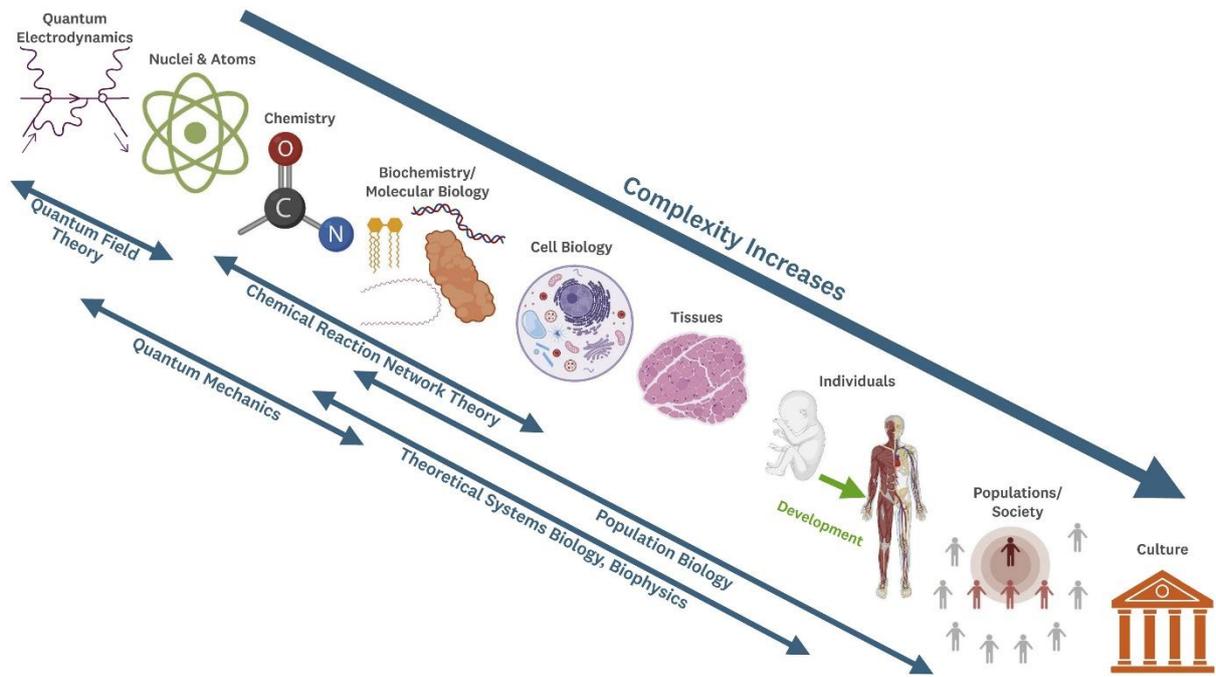

**Figure 1**. Illustration of levels of complexity in science (From Stumpff, 2022 with permission from Elsevier).